\documentclass[aps,pra,twocolumn,showpacs,superscriptaddress,groupedaddress]{revtex4}

\usepackage{graphicx}
\usepackage{amsthm}
\usepackage[toc,page,title,titletoc,header]{appendix}
\usepackage{graphicx}
\usepackage{epstopdf}
\usepackage{CJK}

\theoremstyle{plain}

\theoremstyle{definition}
\newtheorem{defn}{Definition}

\theoremstyle{remark}

\newcommand{\Tr}{\text{Tr}}
\newcommand{\ket}[1]{|#1\rangle}
\newcommand{\bra}[1]{\langle#1|}
\begin{document}
\begin{CJK*}{GB}{gbsn}
\title{Dynamical Invariants of Open Quantum Systems}
\author{S. L. Wu(武松林)}
\email{slwu@dlnu.edu.cn}
\affiliation{School of Physics and Materials Engineering,\\
Dalian Nationalities University, Dalian 116600 China}
\author{X. Y. Zhang(张兴远)}
\affiliation{Center for Quantum Sciences and School of Physics,\\
Northeast Normal University, Changchun 130024, China}
\author{X. X. Yi(衣学喜)}
\email{yixx@nenu.edu.cn}
\affiliation{Center for Quantum Sciences and School of Physics,\\
Northeast Normal University, Changchun 130024, China}

\date{\today}
\begin{abstract}
For a closed quantum system, a dynamical invariant is defined as an
operator whose expectation value is a constant. In this paper, we
extend the concept of  dynamical invariants from closed systems   to
open systems. A dynamical equation for invariants (the dynamical
invariant condition) is derived  for Markovian dynamics. Different
from dynamical invariants of closed quantum systems, the time
evolution of dynamical invariants of open quantum systems is no
longer unitary, and eigenvalues of any invariant are time-dependent in general.
Since any hermitian operator which can fulfill the dynamical invariant
condition is a dynamical invariant, we construct a type of special
dynamical invariants of which a part of the eigenvalues is still
constant. The dynamical invariants in the subspace spanned by these
eigenstates thus evolve unitarily.
\end{abstract}

 \maketitle

\end{CJK*}

\section{introduction}

Dynamical invariants of a  closed system, defined as hermitian
operators  with time-independent expectation values, were proposed
by Lewis half a century ago\cite{lewis}. This theory was designed to
investigate the time evolution of dynamical systems with an
explicitly time-dependent Hamiltonian\cite{lr}. Thereafter, the
dynamical invariants are successfully applied to investigate
 time-dependent problems in quantum mechanics\cite{es} such as the
Berry phase\cite{gp}, and the connection between quantum theory and
classical theory\cite{qcc} was established with the help of these
invariants. Recently, this theory has attracted attention again due
to its application in {\it shortcuts to adiabaticity} to speed up
adiabatic and quantum control processes \cite{chen,chen1}. Such
scheme is also known as the {\it inverse engineering technology}, which has
been widely used in a number of quantum controls and quantum
information processings\cite{qc1,qc2,qc3,qc4}.

Although the theory of  dynamical invariants is  powerful in the
study of closed quantum systems, the definition of dynamical
invariants for open quantum systems is still rarely studied. In this
paper, we generalize the dynamical invariants from closed systems to
open systems. Noticing  that the dynamical invariants are in fact
hermitian operators whose expectation values remain constant in time
evolution\cite{lewis}, we then can define a dynamical invariant for
open systems and derive  the dynamical equation of those invariants
(the dynamical invariant condition) based on this spirit. We should
emphasize   that  the dynamics of open quantum systems can be
described  in many ways, e.g., master
equation\cite{master1,master2,master3}, Kraus
representation\cite{kraus1} and so on. This fact may make the
definition of dynamical invariants for open systems different with
respect to that of closed systems. Dynamical invariants are not unique
even for a given description of open systems, as we show later.

As well known, when a quantum state is initially in a
decoherence-free subspace (DFS), the quantum state would evolve
unitarily \cite{dfs1, dfs2, dfs3}. For a quantum system
governed by a Markovian master equation in the Lindblad form
\begin{eqnarray}
\dot{\rho}(t)&=&-i[H(t),\rho(t)]+\mathcal{L}\rho(t),\nonumber\\
\mathcal{L}\rho(t)&=&\sum_\alpha \left[F_\alpha \rho(t)
F_\alpha^\dag-\frac{1}{2}\{F_\alpha^\dag F_\alpha,
\rho(t)\}\right], \label{tmq}
\end{eqnarray}%
with Lindblad operator $F_\alpha(t)$,  a subspace spanned by
$\mathcal H_{\text{DFS}}:=\{\ket
{\Phi_1},\ket{\Phi_2},...,\ket{\Phi_D}\}$ is a D-dimensional
decoherence free subspace inside  the $N$-dimensional Hilbert space,
if and only if the following conditions are
fulfilled\cite{dfs1,dfs2, dfs3}: (1) All of the bases of $\mathcal
H_{\text{DFS}}$ are the degenerated eigenstates of all Lindblad
operators $F_\alpha(t)$ with common eigenvalues $c_\alpha(t)$, i.e.,
\begin{eqnarray}
F_\alpha(t)\ket{\Phi_j(t)}=c_\alpha(t)\ket{\Phi_j(t)} \label{c1}
\end{eqnarray}
for $\forall j,\alpha$, and (2) $\mathcal H_{ \text{DFS}}$ is
invariant under the following operator
\begin{eqnarray}
H_{\text{eff}}&=&G(t)+ H(t)  \nonumber\\&&+\frac {i}{2} \sum_
\alpha\left(c^*_\alpha(t) F_\alpha(t)-c_\alpha(t)
F^\dag_\alpha(t)\right), \label{c2}
\end{eqnarray}
in which
$$G(t)=i\left(\sum_{j=1}^D \ket{\Phi_j(t)}\bra{\dot \Phi_j(t)}+\sum_{n=1}^{N-D}
\ket{\Phi_n^\bot(t)} \bra{\dot \Phi_n^\bot(t)}\right).$$  Here we
use the notation $\ket{\Phi_n^\bot(t)} (n=1,...,N-D)$  as the
basis of complimentary subspace $\mathcal H^\bot$ of DFS
\cite{tdfs1}.  Actually, the condition (2) requires that any
quantum state in the DFSs must still be a quantum state of  the DFS after
the action of  $H_{\text{eff}}$,  i.e.,
\begin{eqnarray}
\bra{\Phi_n^\bot(t)}H_{\text{eff}}\ket{\Phi_j(t)}=0,\,\forall n,j.\label{hn}
\end{eqnarray}
This is also known as the DFS condition.

Since any hermitian operator which satisfies  the dynamical
invariant condition is a dynamical invariant of the
system\cite{qc1,uu2}, the following questions are interesting:
whether there exists a kind of dynamical invariants whose evolution
is unitary? What are the conditions for  such  dynamical
invariants? In this paper, we will answer those questions by looking
for a sort of dynamical invariants which is block-diagonal. The upper
block evolves unitarily, and the lower block decouples to the upper
block. Our results show that this type of  dynamical invariants
exists, if a DFS can be found  in the dynamics of quantum
systems.

This paper is structured as follows.  We begin by defining a
dynamical invariant  and presenting the dynamical invariant condition
for an open system described by the Markovian master equation in
Sec.\ref{DIO}. After determining the dynamical invariant condition,  in Sec.\ref{DIDFS},
we construct  a type of dynamical invariants in which a part of it that belongs to the decoherence free subspace decouples
to the other part. In Sec.\ref{dephas}, a pure
dephasing model is considered to illustrate  how to construct  the
dynamical invariants   proposed in this paper. Finally, Sec.\ref{conc}
summarizes the results and presents the conclusions.

\section{General formalism}

\subsection{Dynamical Invariants of an Open Quantum System}\label{DIO}

Let us start with the Lewis-Riesenfeld's  dynamical invariant theory
for closed quantum systems\cite{lewis}. A dynamical invariant
$I(t)$ is defined as a time-dependent hermitian operator, which
satisfies the von Neumann-like equation,
\begin{eqnarray}
\frac{\partial I(t)}{\partial t}+i [H(t),I(t)]=0,\label{cinv}
\end{eqnarray}%
where $H(t)$ is a time-dependent Hamiltonian. Obviously, the dynamical
invariants comply with the following properties: (a) The expectation
values of the dynamical invariants are constant. (b) The eigenvalues of a
dynamical invariant are constant, while the eigenstates are
time-dependent. (c) Any time dependent hermitian operator
which satisfies Eq.(\ref{cinv}) is a dynamical invariant for closed
quantum systems. Each dynamical invariant corresponds to a symmetry of
the closed quantum system.

From the definition of dynamical invariants for closed systems,
we find that  the dynamical invariants for  open quantum systems  cannot
be obtained by  generalizing the  Lewis-Riesenfeld's theory
directly. Nevertheless, the essence of the dynamical
invariants is the conservation quantity of quantum systems. Therefore,
we define a dynamical invariant for open
systems as follows,
\begin{defn}
A dynamical invariant $I(t)$ of a quantum open system is a
time-dependent hermitian operator, whose expectation value is a
constant, i.e.,
\begin{eqnarray}
\frac{d}{dt} \langle I(t)\rangle=0.\label{dinvar}
\end{eqnarray}
\end{defn}
On the one hand, when the quantum system is closed, the
dynamical invariants are back to that for closed systems,
i.e., it satisfies  the von Neumann-like equation. Therefore, the
definition we proposed here covers the Lewis's definition
\cite{lewis}. Actually, Eq.(\ref{dinvar}) is nothing but  the
dynamical equation of the invariants. So in the later discussions,
we name the dynamical equation of the invariants as \emph{dynamical
invariant condition}. On the other hand, once the dynamics of an
open quantum system is given, the dynamical invariant condition can be derived.

To give an  explicit dynamical invariant condition, we need to
specify open quantum systems. \textbf{An open
system described by a Lindblad Markovian master equation is a good example, see
Eq.(\ref{tmq}).} By the definition of the dynamical invariants,
the expectation value of a dynamical invariant satisfies
\begin{eqnarray}
\frac{d}{dt} \langle I(t)\rangle=\Tr \{ \dot I(t) \rho(t)+ I(t) \dot\rho(t)\}=0.
\end{eqnarray}%
After the consideration of the dynamical equation  Eq.(\ref{tmq}),
we rewrite the above equation as,
\begin{eqnarray}
&&\Tr \left(\left(\dot I(t)+i[H(t),I(t)]\right.\right.\nonumber\\
&&\left.\left.+\frac{1}{2}\sum_\alpha(2F_\alpha^\dag
I(t)F_\alpha-\{F_\alpha^\dag F_\alpha,
I(t)\})\right)\rho(t)\right)=0.
\end{eqnarray}%
Therefore, the dynamical invariants for open quantum systems must obey
\begin{eqnarray}
\frac{\partial I(t)}{\partial t}&+&i[H(t),I(t)]
\nonumber\\&+&\sum_\alpha ( F_\alpha^\dag
I(t)F_\alpha-\frac{1}{2}\{F_\alpha^\dag F_\alpha,
I(t)\})=0.\label{einv}
\end{eqnarray}
This is the dynamical invariant condition for the  open system
described by Eq.(\ref{tmq}). Different from the dynamical invariants
of closed quantum systems, the evolution of the dynamical invariants
of open quantum systems is not unitary, but it is still
trace-conserving.

Since the dynamical invariants are hermitian as we assumed earlier,
they can be written in spectral decomposition,
\begin{eqnarray}
I(t)=\sum_k \lambda_k\ket{\psi_k(t)}\bra{\psi_k(t)},\label{sd}
\end{eqnarray}
where $\lambda_k$ is the $k$-th eigenvalue of $I(t)$  and
$\ket{\psi_k(t)}$ is the corresponding eigenstate. Because the evolution
of the dynamical invariants is not unitary, the eigenvalues might be
time-dependent. This can be seen by substituting
Eq.(\ref{sd}) into the dynamical invariant condition
Eq.(\ref{einv}). After simple algebra, the evolution of the
eigenvalue $\lambda_k$ follows,
\begin{eqnarray}
\frac{\partial }{\partial t}\lambda _{k}&= &\bra{\psi_k(t)}\frac{\partial}{\partial t}I(t)\ket{\psi_k(t)}.\nonumber\\
&=& \bra{ \psi_{k}(t)}\sum_{\alpha}(\lambda _{k} F_{\alpha}^{\dag } F_{\alpha}-
F_{\alpha}^{\dag }I(t)F_{\alpha})\ket{ \psi _{k}(t)}.\label{eigval}
\end{eqnarray}
A proof of this statement can be found in Appendix \ref{deig}.
Note that the eigenvalues are only affected by the decoherence  process.
One  may wonder whether the dynamical invariants can be designed
properly, such that some of the eigenvalues are time-independent.
For instance, when $\ket{ \psi _{k}(t)}$ is the degenerate
eigenstate of $F_{\alpha}$ ($\forall \alpha$), the corresponding
eigenvalue $\lambda_k$ remains constant. This is possible as will show later,   a type
of dynamical invariants for open quantum systems, in which a part of the
eigenvalues is constant, can be constructed.

\subsection{Decoherence Free Subspace and Dynamical Invariants}\label{DIDFS}

Generally speaking, any hermitian operator fulfilling
Eq.(\ref{einv}) is a dynamical invariant of open quantum systems.
In this subsection,  we propose a type of dynamical invariants  with
part of its eigenvalues being constant. As discussed earlier, the
eigenvalues would be constant, if the corresponding eigenstates are
in DFSs. Thus, we may construct the dynamical invariants in the
form,
\begin{eqnarray}
I(t)=\left(
       \begin{array}{cc}
         I^D(t) & 0 \\
         0 & I^C(t) \\
       \end{array}
     \right).
 \label{dinv}
\end{eqnarray}
In Eq.(\ref{dinv}), $I^D(t)$ is the part of this dynamical invariant
belonging  to the DFS, i.e.,
\begin{eqnarray}
I^D(t)=\sum_{i,j} I^D_{ij}(t)\ket{\Phi_i(t)}\bra{\Phi_j(t)};
\end{eqnarray}
$I^C(t)$ is the other part of the dynamical invariants  belonging
to the complementary subspace, which can be decomposed by the bases
of the complementary subspace,
\begin{eqnarray}
I^C(t)=\sum_{mn} I^C_{mn}(t)\ket{\Phi_m^\bot(t)}\bra{\Phi_n^\bot(t)}.
\end{eqnarray}
There are three key features for this kind of dynamical invariants:  (1) the
eigenvalues of $I^D(t)$ are constant; (2) $I^D$ decouples to $I^C$;
(3) $I^D$ and $I^C$ can both be used as the dynamical invariants of
open quantum systems independently.

The Lindblad operators can also be rewritten in block form by the
same merit as in Ref.\cite{lidar},
\begin{eqnarray}
F_\alpha=\left(
    \begin{array}{cc}
         c_\alpha\mathcal{I}^D & A_\alpha \\
    0 & B_\alpha \\
    \end{array}
  \right)\label{F}
\end{eqnarray}
where $\mathcal I^D$ is the identity operator in the DFS. The  upper
(lower) block acts entirely inside $\mathcal H_{\text{DFS}}$
$(\mathcal H^\bot)$; the off-diagonal block $A_\alpha$ mixes
$\mathcal H_{\text{DFS}}$ and $\mathcal H^\bot$. The presence of
$A_\alpha$ is permitted since the DFS condition Eq.(\ref{c1}) that
gives no information about the invariant of $\mathcal H_{
\text{DFS}}$ acted by $H_{\text{eff}}$. Likewise, it is convenient
to rewrite the Hamiltonian $H$ and $G$ as
\begin{eqnarray}
H=\left(
    \begin{array}{cc}
         H^D & H^N \\
    H^{N\dag} & H^C \\
    \end{array}
  \right)\label{H}
\end{eqnarray}
and
\begin{eqnarray}
G=\left(
    \begin{array}{cc}
         G^D & G^N \\
    G^{N\dag} & G^C \\
    \end{array}
  \right)
\end{eqnarray}
with the blocks on the diagonal corresponding once again to the
operators restricted to $\mathcal H_{ \text{DFS}}$ and $\mathcal
H^\bot$. Here we do not restrict the discussion to whether the DFS
is time dependent (t-DFS) or not (traditional DFS). When the DFS is
time independent, the basis $\ket{\Phi_j}$ and $\ket{\Phi_n^\bot}$
$(\forall j, n)$ are constant vectors, i.e., $G=0$.

After putting Eq.(\ref{dinv}) into the
general dynamical invariant condition Eq.(\ref{einv}), the dynamical
invariants $I^D$ and $I^C$ are governed by the following equations,
\begin{eqnarray}
&\dot I^D&+i[G^D+H^D,I^D]=0,\label{id}\\
&\dot I^C&+i[G^C+H^C+\frac{i}{2}\sum_\alpha A_\alpha^\dag A_\alpha,I^C]
+\sum_\alpha A_\alpha^\dag I^DA_\alpha\nonumber\\&&-\frac{1}{2}
\sum_\alpha(\{B_\alpha^\dag B_\alpha,I^C\}-2B_\alpha^\dag I^C B_\alpha)=0,\label{ic}\\
&I^D&(i(G^N+H^N)-\sum_\alpha\frac{c_\alpha^*}{2}A_\alpha)\nonumber\\
&&-(i(G^N+H^N)-\sum_\alpha \frac{c_\alpha^*}{2}A_\alpha)I^C=0.\label{decoupling}
\end{eqnarray}
Here we suppose that the bases of DFSs and complementary subspaces
are time-dependent. The operator $G$ results from  the time
dependent bases of the DFS and its complementary subspace.

On the one hand, Eq.(\ref{decoupling}) is nothing
but the decoupling condition of $I^D$ and $I^C$.  If the decoupling
condition fulfills, the dynamical invariants we proposed would  exist
and the off-diagonal element of Eq.(\ref{dinv}) would
vanish  permanently. It is easy to check that the decoupling condition is
satisfied when the decoherence free subspace emerges: As mentioned
in the DFS condition, $\mathcal H_{\text{DFS}}$ is invariant under
the operator $H_{\text{eff}}$. In other words, the vanishing
off-diagonal element of $H_{\text{eff}}$ guarantees the invariant of
$\mathcal H_{\text{DFS}}$, which is presented as Eq.(\ref{hn}).
Considering block form of the invariants, the vanishing off-diagonal element can
be written as
\begin{eqnarray}
i(G^N+H^N)-\sum_\alpha \frac{c_\alpha^*}{2}A_\alpha=0.
\end{eqnarray}
After taking the DFS condition into Eq.(\ref{decoupling}), the  decoupling
condition is always held. Therefore, the decoupling condition is
nothing but the DFS condition.

On the other hand, the evolution of $I^D$ $(I^C)$ is  governed by
Eq.(\ref{id}) (Eq.(\ref{ic})). Firstly, we may observe from
Eq.(\ref{id}) that the evolution of $I^D$ is unitary and not related
to $I^C$, which is the very dynamical invariant condition for the
closed quantum system (Eq.(\ref{cinv})). In other words, the
evolution of $I^D$ is decoherence free, hence we would like to
name it \emph{decoherence free dynamical invariants}. As a result,
as soon as Eq.(\ref{c1}) is placed into Eq.(\ref{eigval}), we can
immediately find that $\lambda_j^D$ is constant due to $\partial
\lambda_j^D/\partial t=0$, where $\lambda_j^D$ is the $j$-th
eigenvalue of $I^D$. As mentioned above, the eigenstates associated
with the constant eigenvalues construct a basis set of the DFS. Therefore,
it is easy to design $I^D$ in accordance with the
common eigenstates of the Lindblad operators. Secondly, the
evolution of the dynamical invariants $I^C$ is governed by Eq.(\ref{ic}).
The dynamics of $I^C$ is similar to the Markovian-Lindblad master
equation with the Lindblad operators $B_\alpha$ and the Hamiltonian
$G^C+H^C-i\sum_\alpha A_\alpha^\dag A_\alpha/2$ except for an
extra term $\sum_\alpha A_\alpha^\dag I^DA_\alpha$. The
extra term determines that the evolution of $I^C$ is not closed, but
is impacted by $I^D$. In consideration of Eq.(\ref{F}), the
eigenvalues of $I^C$ satisfy
\begin{eqnarray}
\frac{\partial}{\partial
t}\lambda_n^C&=&\sum_j(\lambda_n^C-\lambda_j^D)
|\bra{\psi_j}A_\alpha\ket{\psi_n^\bot}|^2\nonumber\\ &&+\sum_m
(\lambda_n^C-\lambda_m^C)|\bra{\psi_m^\bot}B_\alpha\ket{\psi_n^\bot}|^2.
\end{eqnarray}
Note that all of the eigenvalues $\lambda_n^C$ are time-dependent.
Furthermore, the evolutions of $\lambda_n^C$ are not
self-determined, but affected by the eigenvalues of $I^D$.

\section{Dynamical Invariants for a Dephasing System}\label{dephas}

As an example, we consider a quantum system consisting of $n$
physical qubits, which interacts collectively with a dephasing
environment \cite{tong}. The Hamiltonian of the quantum system can
be written as,
\begin{eqnarray}
H_0=\sum_{i<j}(g_{ij}^x(t) O_{ij}^x+B_{i}^z(t) O_{i}^z),
\end{eqnarray}
in which $g_{ij}^x(t)$ and $B_{i}^z(t)$ denote controllable
coupling strengths, and
\begin{eqnarray}
&&O_{ij}^x=\frac{\sigma_i^x\sigma_j^x+\sigma_i^y\sigma_j^y}{2},
\nonumber\\ &&O_{i}^z=\sum_{i=1}^n\frac{\sigma_i^z }{2}.
\end{eqnarray}
The operator $O_{ij}^x$ is the $XY$ interaction between the $i$-th
qubit and the $j$-th qubit, where $\sigma^x_i$ and $\sigma^y_i$ are
Pauli operators for the $i$-th qubit. This Hamiltonian can be used
to describe a number  of quantum systems, such as trapped ions and
quantum dots \cite{dp1,dp2,dp3,dp4}. The source of   decoherence in
the quantum system considered here is a dephasing environment. The
interaction between the quantum system and the environment can be
described by,
\begin{eqnarray}
H_I=\sum_j \sigma^z_j \otimes \hat{B},
\end{eqnarray}
where $\hat{B}$ is the environment operator. The reduced dynamics of the
quantum system is described by the following master equation
\cite{dpm},
\begin{eqnarray}
\dot \rho(t)=-i[H_0,\rho(t)]+\mathcal L(\rho).
\end{eqnarray}
Here the dephasing is characterized by    $$\mathcal
L(\bullet)=\gamma(F\bullet F-\bullet),$$ with
\begin{eqnarray}
F=\sum_j\sigma_j^z.\label{dF}
\end{eqnarray}
It is well known that this master equation is a Lindblad  mater
equation and $F$ is the corresponding Lindblad operator. The
symmetry of the master equation implies that there exists a DFS. Our
aim is to find the dynamical invariants of this open quantum system
which are block-diagonal as in Eq.(\ref{dinv}).

To simplify the problem, we assume that the quantum system consists
of two qubits, i.e., $N=2$. The case with qubits more than two can
be studied by the same process as that presented below. For a
two-qubit system, the corresponding DFS is spanned by
$\{\ket{01},\ket{10}\}$ which are the degenerate eigenstates of $F$
with zero eigenvalue. The bases of the complementary subspace are
chosen as $\{\ket{00},\ket{11}\}$. According to Eq.(\ref{dF}), the
Lindblad operator $F$ can be written in block from as in
Eq.(\ref{F}) with $c=0$, and
\begin{eqnarray}
A=\left(
           \begin{array}{cc}
             0 & 0 \\
             0 & 0 \\
           \end{array}
         \right),
B=\left(
           \begin{array}{cc}
             -2 & 0 \\
             0 & 2 \\
           \end{array}
         \right).
\end{eqnarray}
The Hamiltonian $H_0$ also has a matrix  representation as Eq.(\ref{H}) with
\begin{eqnarray}
&&H^D=\left(
           \begin{array}{cc}
             0 & g^x_{12}(t) \\
             g^x_{12}(t) & 0 \\
           \end{array}
         \right),\\
&&H^C=\left(
           \begin{array}{cc}
             -B^z(t) & 0 \\
             0 & B^z(t) \\
           \end{array}
         \right),\\
&&H^N=\left(
           \begin{array}{cc}
             0 & 0 \\
             0 & 0 \\
           \end{array}
         \right).
\end{eqnarray}

In order to verify the dynamical invariants $I(t)$ which can be
expressed as Eq.(\ref{dinv}), it is necessary to check if the
decoupling condition (Eq.({\ref{decoupling}})) is satisfied. Since
the DFS is time-independent, $G$ always vanishes in
Eq.({\ref{decoupling}}). Moreover, we have already shown that
the off-diagonal elements of both the Lindblad operator and
 the Hamiltonian are trivial. Namely, $A=H^N=\textbf{0}$,
where $\textbf{0}$ denotes the zero matrix (all elements of the
matrix are zero). Therefore, the dynamical invariants $I(t)$ can be
written in the form of Eq.(\ref{dinv}).

Next, let us derive  the upper and lower nonzero  block elements of
$I(t)$, i.e., $I^D$ and $I^C$. The evolution of $I^D$ is unitary,
which is governed by Eq.(\ref{id}),
\begin{eqnarray}
\dot I^D&+i[H^D(t),I^D]=0.
\end{eqnarray}
The Hamiltonian $H^D(t)$ can be rewritten in the form of the pauli
matrixes,
\begin{eqnarray}
H^D(t)=g_{12}^x(t) \sigma_x.
\end{eqnarray}
By SU(2) algebra of the pauli matrixes, a dynamical invariant that
is hermitian  can also be expended by the pauli matrixes,
\begin{eqnarray}
I^D=x^D\sigma_x+y^D\sigma_y+z^D\sigma_z.
\end{eqnarray}
with  coefficients $x^D$, $y^D$, $z^D$ to be derived.  Substituting
$H^D$ and $I^D$ into Eq.(\ref{id}), the equations of motion for
those expansion coefficients are given by,
\begin{eqnarray}
&&\dot{x^D}=0,\nonumber\\
&&\dot{y^D}=2g_{12}^x(t) z^D,\nonumber\\
&&\dot{z^D}=-2g_{12}^x(t) y^D.\label{ddg}
\end{eqnarray}
The solutions of these equations subject to the given initial conditions
$x^D_0$, $y^D_0$, $z^D_0$ can be easily obtained,
\begin{eqnarray}
&&x^D(t)=x^D_0,\label{xd}\\
&&y^D(t)=y^D_0\cos\left(2\Lambda(t)\right)+z^D_0\sin\left(2\Lambda(t)\right),\label{yd}\\
&&z^D(t)=z^D_0\cos\left(2\Lambda(t)\right)-y^D_0\sin\left(2\Lambda(t)\right),\label{zd}
\end{eqnarray}
where $\Lambda(t)=\int_0^t g_{12}^x(t') dt'$.  The process  of
obtaining these solution is shown in Appendix \ref{solution}.
Consequently we confirm all of the parameters in $I^D$. When the
initial conditions for those parameters are given, the dynamical
invariants $I^D$ can be constructed.

For the lower diagonal-block of the dynamical invariants
Eq.(\ref{F}), the evolution is governed by Eq.(\ref{ic}). Because of
the vanishing off-diagonal block of $F$, i.e., $A=\bf{0}$, the
dynamical invariants $I^C$ can be simplified into,
\begin{eqnarray}
\dot I^C+i[H^C,I^C]-\frac{\gamma}{2}\left(\{B^\dag B,I^C\}-2B^\dag I^C B\right)=0,\nonumber\\\label{icd}
\end{eqnarray}
where
\begin{eqnarray}
H^C=-B^z(t) \sigma_z, B=-2 \sigma_z.\nonumber
\end{eqnarray}
As an hermitian operator, $I^C$ can be written in the following
form,
\begin{eqnarray}
I^C=x^C\sigma_x+y^C\sigma_y+z^C\sigma_z.\label{expc}
\end{eqnarray}
where $x^C$, $y^C$ and $z^C$ are real time-dependent parameters.
Substituting  $I^C$ in Eq.(\ref{icd}), the time development of those
parameters is governed by,
\begin{eqnarray}
&&\dot{x^C}=-2B^z(t)y^C+8\gamma x^C,\\
&&\dot{y^C}=2B^z(t)x^C+8\gamma y^C,\\
&&\dot{z^C}=0.\label{dcg}
\end{eqnarray}
On the one side, as we expected, $z^C$ is not affected by the dephasing
environment. On the other side, $x^C$ and $y^C$ satisfy a set of
coupled differential equations. Here, we may introduce new
parameters as follows,
\begin{eqnarray}
&&x'(t)=x^C(t)\exp(-8\gamma t),\nonumber\\ &&y'(t)=y^C(t)\exp(-8\gamma t),
\end{eqnarray}
which fulfill
\begin{eqnarray}
&&\dot{x'}=-2B^z(t)y',\\
&&\dot{y'}=2B^z(t)x'.\label{x'}
\end{eqnarray}
The new parameters $x'$ and $y'$ satisfy the following second order
differential equations,
\begin{eqnarray}
&&\ddot{x'}-\frac{\dot B^z(t)}{B^z(t)}\dot x'+4B^{z2}(t)x'=0,\\
&&\ddot{y'}-\frac{\dot B^z(t)}{B^z(t)}\dot y'+4B^{z2}(t)y'=0.\label{xy}
\end{eqnarray}
And the solutions can be expressed as,
\begin{eqnarray}
x'(t)&=&x^C_0\cos\left(2\Theta(t)\right)-y^C_0\sin\left(2\Theta(t)\right),\nonumber\\
y'(t)&=&y^C_0\cos\left(2\Theta(t)\right)+x^C_0\sin\left(2\Theta(t)\right),
\end{eqnarray}
where $\Theta(t)=\int_0^t B^z(t') dt'$. Taking the solutions of $x'$ and $y'$ into
Eq.(\ref{x'}) and considering the initial condition of $I^C$, the
solution of $I^C$ can be written as a function of
\begin{eqnarray}
&&x^C(t)=\left(x^C_0\cos\left(2\Theta(t)\right)-y^C_0\sin\left(2\Theta(t)\right)\right)\exp(8\gamma t),\nonumber\\
&&y^C(t)=\left(y^C_0\cos\left(2\Theta(t)\right)+x^C_0\sin\left(2\Theta(t)\right)\right)\exp(8\gamma t),\nonumber\\
&&z^C(t)=z^C_0,\label{zc}
\end{eqnarray}
where $x^C_0$, $y^C_0$, $z^C_0$ are the initial values of $x^C$, $y^C$, $z^C$.

With the  assistance of the analytic solution of $I(t)$,  we can
further discuss the eigenvalues and the eigenstates . The eigenvalues
of $I^D$ and $I^C$ have the same structure,
\begin{eqnarray}
\lambda^o_\pm=\pm\sqrt{x^{o2}+y^{o2}+z^{o2}},\label{ice}
\end{eqnarray}
which is associated with the eigenstate as follows,
\begin{eqnarray}
\ket{\psi^o_\pm}=\left(
                  \begin{array}{c}
                    (\lambda^o_\pm+z^o)/p^o_\pm \\
                    (x^o-i y^o)/p^o_\pm \\
                  \end{array}
                \right)
,
\end{eqnarray}
where $p^o_\pm=\sqrt{2\lambda^o_\pm(\lambda^o_\pm+z^o)}$  is the
normalized coefficient, and $o\in (D,C)$. Examining Eq.(\ref{xd}),
Eq.(\ref{yd}) and Eq.(\ref{zd}), we find that the eigenvalues of
$I^D$ are constant, even though $y^D$ and $z^D$ are time-dependent,
whereas the corresponding eigenstates are time-dependent. This
result is very similar with the  that for a dynamical invariant of
closed quantum systems, which reduces  to the unitary dynamics of
$I^D$. Besides, $I^C$ is affected by the dephasing noise. The
eigenvalues of it are not time-independent anymore, but depend on
the decoherence process. After putting  the solution of $I^C$ into
Eq.(\ref{ice}), we obtain the following eigenvalues
\begin{eqnarray}
\lambda^C_\pm(t)=\pm\sqrt{(x^{C2}_0+y^{C2}_0)\exp(16\gamma t)+z^{C2}_0}.
\end{eqnarray}
Note $\exp(16\gamma t)$ in the solution.

The eigenstates of $I^C$ can also be written as a function of time,
\begin{eqnarray}
\ket{\psi^C_\pm(t)}=\left(
                  \begin{array}{c}
                    \frac{(\lambda^C_\pm(t)+z^C_0)}{p^C_\pm(t)} \\
                    \frac{(x^C_0-i y^C_0)}{p^C_\pm(t)}\exp(-2 i \Theta(t)+8\gamma t) \\
                  \end{array}
                \right),
\end{eqnarray}
in which the normalized constant is also  time-dependent
$p^C_\pm=\sqrt{2\lambda^C_\pm(t) (\lambda^C _\pm(t)+z^C_0)}$. If we
set $z^C_0=0$, the eigenvalues and eigenstates can be further
simplified to,
\begin{eqnarray}
\lambda^C_\pm(t)=\pm\sqrt{(x^{C2}+y^{C2})}\exp(8\gamma t),
\end{eqnarray}
and
\begin{eqnarray}
\ket{\psi^C_\pm(t)}=\left(
                  \begin{array}{c}
                    \pm\sqrt{2}/2 \\
                    \sqrt{2}/2\exp(-2 i \Theta(t)) \\
                  \end{array}
                \right),
\end{eqnarray}
In this special case, the time-dependence  of the eigenvalues  is
dominated by the decoherence process, but the eigenstates can be
only determined  by the external parameters of the open quantum
system.

\section{conclusion}\label{conc}

This work answers the question regarding the   condition required
for a dynamical invariant of an open  quantum system. According to
the definition, the expectation values of  dynamical invariants are
constant, thus we derive the condition for these invariants. We
further construct a type of dynamical invariants in an explicit form
when the open system obeys the Lindblad master equation (as shown in
Eq.(\ref{dinv})). Furthermore,  a relation between the dynamical
invariants and the decoherence free subspaces is established.

Practically speaking, the dynamical invariants of an open quantum
system provide us with both an intuitive physical framework and a
set of tools to understand and manipulate quantum states, especially
for quantum systems with time-dependent Hamiltonian\cite{tham}
and described by Markovian master equation\cite{nmar}. These tools are
proven useful in theoretical formulations of the decoherence-free
subspaces and further experimental developments of quantum
control\cite{chen1,qcc2}.

{\it Note added: When we finished this Manuscript, we had noticed that in
a recent paper, arXiv:1510.00518, the authors developed a method to
find the dynamical invariants for open systems based on the
non-Markovian quantum state diffusion equation.}

\appendix

\section{The Evolution of the Eigenvalues of Dynamical
Invariants}\label{deig}

In Appendix A, we derive the evolution equation for the eigenvalues
of the dynamical invariants. Obviously, the eigenvalue can be obtained by
\begin{eqnarray}
 \lambda_k(t)=\bra{\psi_k(t)} I(t) \ket{\psi_k(t)}.
\end{eqnarray}
By differentiating the eigenvalue with respect to time, we obtain
\begin{eqnarray}
\dot \lambda_k(t)&=&\bra{\dot\psi_k(t)} I(t) \ket{\psi_k(t)}+\bra{\psi_k(t)} I(t) \ket{\dot\psi_k(t)}\nonumber\\
&+&\bra{\psi_k(t)} \dot I(t) \ket{\psi_k(t)}.
\end{eqnarray}
By considering that $I(t)\ket{\psi_k(t)}=\lambda_k\ket{\psi_k(t)}$
and $\frac{ \partial} {\partial
t}(\langle\psi_k(t)\ket{\psi_k(t)})=0$, we find
\begin{eqnarray}
\dot \lambda_k(t)=\bra{\psi_k(t)} \dot I(t) \ket{\psi_k(t)}.
\end{eqnarray}
Substituting  Eq.(\ref{einv}) into the above equation, the evolution of
the eigenvalue can be rewritten as
\begin{eqnarray}
\dot \lambda_k(t)=-\sum_\alpha \bra{\psi_k(t)} ( F_\alpha^\dag
I(t)F_\alpha-\frac{1}{2}\{F_\alpha^\dag F_\alpha,
I(t)\}) \ket{\psi_k(t)}.\nonumber
\end{eqnarray}
We immediately obtain the equation   in the maintext of
Sec.\ref{DIO}, i.e.,
\begin{eqnarray}
\dot \lambda_k(t)=\sum_\alpha \bra{\psi_k(t)} (\lambda_k F_\alpha^\dag F_\alpha -F_\alpha^\dag I(t)F_\alpha) \ket{\psi_k(t)}.
\end{eqnarray}
Here that $\ket{\psi_k(t)}$ is the eigenstate of $I(t)$ with the
eigenvalue $\lambda_k$ being used.

\section{The solution of time-dependent second order differential equation} \label{solution}

In Appendix B, we present the details to solve the time-dependent
second order differential equation used in Sec.\ref{dephas}. As
shown in Eq.(\ref{ddg}), the parameters $y^D$ and $z^D$ satisfy the
following second order differential equation,
\begin{eqnarray}
&&\ddot{y^D}-\frac{\dot g_{12}^x(t) }{g_{12}^x(t) }\dot y^D+4g_{12}^x(t)  y^D=0,\\
&&\ddot{z^D}-\frac{\dot g_{12}^x(t) }{g_{12}^x(t) }\dot z^D+4g_{12}^x(t) z^D=0. \label{zdd}
\end{eqnarray}
We can find that $y^D$ and $z^D$ satisfy similar differential
equations, so do the parameters $x'$ and $y'$ in Eq.(\ref{xy}).
Without loss of generality, we take Eq.(\ref{zdd}) as an example. In order to
obtain the solution of Eq.(\ref{zdd}), we introduce a new parameter
$u$, which can be written as
\begin{eqnarray}
u=-\frac{\dot z^D}{g_{12}^x(t) z^D}.\label{ud}
\end{eqnarray}
Taking the first and second order time  derivative of $z^D$ into
Eq.(\ref{zdd}), we immediately obtain a first order differential
equation about the new parameter $u(t)$,
\begin{eqnarray}
\dot u(t)=g_{12}^x(t)(u(t)^2+4).
\end{eqnarray}
By separating the variables, we rewrite the above equation as,
\begin{eqnarray}
\frac{d u(t)}{u(t)^2+4}=g_{12}^x(t)dt.
\end{eqnarray}
Thus the solution of $u$ can be obtained,
\begin{eqnarray}
u(t)=2\tan(2\Lambda(t)+A).
\end{eqnarray}
Here $A=\arctan(\dot z^D(0)/2g_{12}^x(0)z^D(0))$. After we apply the
solution of $u(t)$ into Eq.(\ref{ud}), the solution of $z^D$ can
finally be determined,
\begin{eqnarray}
z^D(t)=z^D(0)\cos(2\Lambda(t)+A).
\end{eqnarray}
By considering $\dot z^D(0)=-2g_{12}^x(0)y^D(0)$, we rewrite the solution of Eq.(\ref{zdd}) as
\begin{eqnarray}
z^D(t)=z^D(0)\cos\left(2\Lambda(t)\right)-y^D(0)\sin\left(2\Lambda(t)\right),
\end{eqnarray}
which is Eq.(\ref{zd}).

\end{document}